\documentclass[a4paper,twocolumn,11pt,accepted=2026-02-09]{quantumarticle}
\pdfoutput=1
\usepackage[utf8]{inputenc}
\usepackage[english]{babel}
\usepackage[T1]{fontenc}
\usepackage{amsmath}
\usepackage[numbers,sort,compress]{natbib}
\usepackage{xcolor}
\definecolor{darkblue}{RGB}{0,0,150}
\definecolor{nightblue}{RGB}{0,0,100}
\usepackage[
  colorlinks,
  citecolor=darkblue,
  linkcolor=darkblue,
  urlcolor=nightblue
]{hyperref}

\usepackage{graphicx}
\usepackage{mathtools}
\usepackage{bm}
\usepackage{bbm}
\usepackage{amssymb}
\usepackage{booktabs}
\usepackage[babel,kerning=true,spacing=true]{microtype}
\usepackage{subfigure}
\usepackage{pifont}
\usepackage{soul}
\usepackage{float}
\usepackage{placeins}

\usepackage{amssymb}
\usepackage{pifont}

\begin{document}

\title{Increasing the distance of topological codes with time vortex defects}
    
\author{Gilad Kishony}
\affiliation{Department of Condensed Matter Physics, Weizmann Institute of Science, Rehovot, 76100, Israel}
\author{Erez Berg}
\affiliation{Department of Condensed Matter Physics, Weizmann Institute of Science, Rehovot, 76100, Israel}

\begin{abstract}
We propose modifying topological quantum error correcting codes by incorporating space-time defects, termed ``time vortices,'' to reduce the number of physical qubits required to achieve a desired logical error rate. A time vortex is inserted by adding a spatially varying delay to the periodic measurement sequence defining the code such that the delay accumulated on a homologically non-trivial cycle is an integer multiple of the period. We analyze this construction within the framework of the Floquet color code and optimize the embedding of the code on a torus along with the choice of the number of time vortices inserted in each direction. Asymptotically, the vortexed code requires less than half the number of qubits as the vortex-free code to reach a given code distance. We benchmark the performance of the vortexed Floquet color code by Monte Carlo simulations with a circuit-level noise model and demonstrate that the smallest vortexed code (with $30$ qubits) outperforms the vortex-free code with $42$ qubits. 

\end{abstract}

\maketitle

\section{Introduction}
\label{sec: intro}

Quantum error correction is key to enabling scalable quantum computing, as it protects quantum information from errors induced by noise and operational imperfections. Topological quantum codes \cite{dennis2002surfacecode,
kitaev2003toriccode, Fowler2012surfacecode, fujii2015quantumcomputationtopologicalcodes, Albuquerque2022} are among the most promising frameworks for error correction due to their reliance on local connectivity, often requiring only two-dimensional (2D) hardware layouts. In these codes, logical qubits are protected by enforcing the eigenvalues of a set of commuting stabilizer operators defined on local patches of qubits arranged in a lattice embedded in a 2D manifold. These local checks allow for the detection of errors as long as the errors do not form homologically non-trivial paths on the lattice—paths whose minimal length scales with the system size. Consequently, as the system’s size increases, the logical qubits gain enhanced protection from errors. In practice, the stabilizers of these codes are measured periodically and the measurement results (syndromes) are cross-referenced over time to overcome measurement errors and to avoid accumulation of errors over time.

Recently, a generalization of topological codes known as Floquet codes has been proposed \cite{Hastings2021dynamically}. While conventional ``static'' topological codes are defined through a set of commuting stabilizers, Floquet codes employ a periodic measurement sequence where the order of measurements is crucial as these do not generally commute with one another. A significant advantage of Floquet codes is that they require the measurement of only low-weight checks at each step. In particular, certain constructions involve only two-qubit parity measurements, which are native operations on certain types of hardware. By taking products of these measurement outcomes over time, it becomes possible to indirectly infer higher-weight stabilizer operators of an underlying static code. If no intermediate checks anti-commute with a particular stabilizer between two consecutive measurements, the product of these two measurements forms a ``detection cell'' in space-time. A detection cell yielding a non-trivial syndrome indicates that an error has occurred.

Floquet codes have been studied extensively since the development of the honeycomb Floquet code \cite{Haah2022boundarieshoneycomb} which measures 6-body stabilizers using 2-body checks. In \cite{Haah2022boundarieshoneycomb, Gidney2021faulttolerant, Gidney2022benchmarkingplanar, planar_with_majoranaPaetznick2023} the realization of boundaries in the honeycomb Floquet code needed for a planar architecture was simplified and its performance was benchmarked for circuit-level noise. Later, Refs. \cite{WithoutSubsystemCodesDavydova2023, Bombin2024unifyingflavorsof, Anyons_color_code_Kesselring2024} constructed a Calderbank-Shor-Steane (CSS) variation called the Floquet color code which was shown to have a competitive threshold. Refs.~\cite{ellison2023floquetcodestwist, twist_defect_networks_Sullivan2023} considered encoding logical qubits using twist defects in Floquet codes. 


In this work we suggest 
introducing ``time vortex'' defects 
as a means of enhancing a static or Floquet topological code's properties. These defects are topological in nature and involve a temporal lag in the measurement sequence, which varies spatially and winds around the
defect core by an integer multiple of the period. We find that this spatial-temporal modification alters the paths through which undetectable errors can form. By shearing the space-time lattice of detector cells, time vortex defects can enhance protection from logical errors by forcing undetectable error paths to become longer in the deformed topology. This construction is inspired by Ref.  \cite{kishony2024topologicalexcitationstimevortices}, in which a time vortex defect is introduced in a unitary Floquet setting.

\begin{figure}[H]
    \centering
    \includegraphics[width=1\columnwidth]{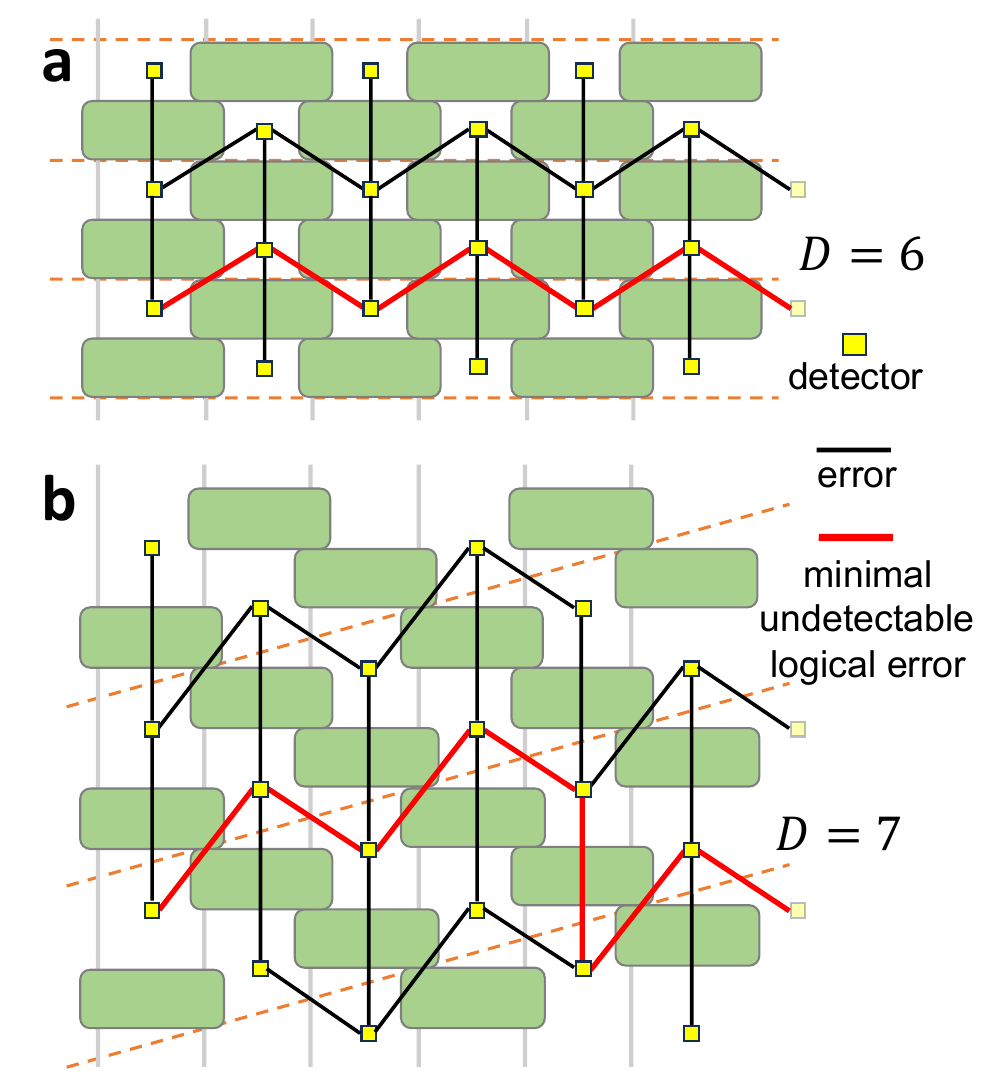}
    \caption{Inserting a time vortex into the measurement sequence of the repetition code with periodic boundary conditions. \textbf{(a)} Without the time vortex, native two-body Pauli $ZZ$ measurements (green) are applied in a ``brick wall'' circuit. 
    The height of the measurement boxes represents the minimal time required to perform a measurement. 
    The matching graph of the code is shown for a phenomenological noise model. Detectors given by the product of two consecutive parity measurements are drawn as yellow square markers. Errors represented by edges trigger the detectors at their endpoints. Vertical (time-like) edges are measurement errors while other (space-like) edges correspond to bit flips between rounds of syndrome measurement (along the dashed orange lines). The code distance $D$ is determined by the length of the minimal non-trivial homotopy (red edges). \textbf{(b)} A single time vortex is inserted by introducing delays between consecutive measurements. The distance is increased by $1$ as compared with the vortex-free code. This comes at the expense of an increased circuit depth.
    }
    \label{fig: matching graph cartoon}
\end{figure}

Our construction is most easily understood by considering the simple repetition code defined by the Pauli stabilizers $Z_iZ_{i+1}$ with $N$ qubits $i\in\{1,\dots,N\}$ and periodic boundary conditions. Fig.~\ref{fig: matching graph cartoon}(a) illustrates the periodic circuit used to measure the syndromes of this code using native two-body Pauli $ZZ$ measurements and the corresponding matching graph (the relationship between errors and detectors in space-time), assuming a phenomenological noise model including measurement errors and single qubit errors between each complete round of measurements (dashed orange lines in Fig.~\ref{fig: matching graph cartoon}). The path along the red edges in the matching graph in Fig.~\ref{fig: matching graph cartoon}(a) is a minimal weight undetectable error. The insertion of a single time vortex is drawn in Fig.~\ref{fig: matching graph cartoon}(b). The same path on the matching graph in this case is no longer a cycle, i.e. it becomes a detectable error. The minimal weight of an undetectable error, or the code distance, has increased by $1$. 

We note that in this example, the increase in the code distance due to the introduction of a time vortex depends on the particular noise model used: if single qubit errors were allowed between any two measurements (rather than between complete measurement rounds), the time vortex would not increase the distance. 
In Appendix \ref{app: rep code with ancillas} we show that the time vortex increases the distance in a more realistic setting, without the use of native two-body Pauli measurements, using auxiliary qubits to measure the syndromes of this code.

Within the Floquet color code framework, we optimize the way in which we embed the code on the torus and the number of time vortices introduced through each hole. By introducing a number of vortices proportional to the linear size of the system through the holes of the torus, the code distance increases compared to the optimal distance achievable with no vortices. Specifically, for a realistic circuit-level noise model, the distance in the vortexed Floquet color code becomes approximately 1.46 times the distance in the vortex-free configuration in the limit of a large number of qubits. This leads to a stronger suppression of the error rate with system size. Using Monte Carlo simulations, we demonstrate the enhanced performance of the vortexed Floquet color code, positioning it as a strong contender among state-of-the-art quantum error correction techniques.


We note that inserting time vortices into a tightly packed quantum circuit increases its depth (while the number of gates remains the same). This leads to a tradeoff between the beneficial increase in code distance on the one hand and increased duration and idling noise on the other. Vortexing becomes useful if there is a low error rate for idle qubits compared to gate infidelities. This is discussed in Appendix \ref{app: idle}.

This paper is organized as follows. In Sec.~\ref{sec: floquet color code} we review the Floquet color code. In Sec.~\ref{sec: inserting time vortices} we introduce the time vortex construction. 
In Sec.~\ref{sec:optimal} we optimize the embedding of the Floquet color code on the torus both with and without time vortices and demonstrate that a larger distance can be achieved in the former case. Sec.~\ref{sec: simulations} includes numerical simulations of the performance of the optimal codes found in Sec.~\ref{sec:optimal} when subjected to circuit-level noise. Finally, in Sec.~\ref{sec: toric code} we explore the time vortex construction in the case of the toric code.

\section{The Floquet color code}
\label{sec: floquet color code}

The Floquet color code \cite{WithoutSubsystemCodesDavydova2023, Bombin2024unifyingflavorsof, Anyons_color_code_Kesselring2024} is defined by a sequence of 2-body Pauli measurements on the bonds of a honeycomb lattice with a qubit located on each vertex. The set of all checks consists of Pauli $XX$ and Pauli $ZZ$ on all bonds of the lattice; the code is of the CSS family. These 2-body checks are designed in order to infer 6-body stabilizer operators inherited from the parent color code, namely products $X^{\otimes 6}$ and $Z^{\otimes 6}$ on hexagonal plaquettes of the lattice. Specifically, the product of three $X$ checks on non-overlapping bonds around a plaquette is equal to the $X$ stabilizer, and the same is true for Pauli $Z$s.

\begin{figure}
    \centering\includegraphics[ trim={0.0cm 0.0cm 0.0cm 0.0cm},clip,width=\columnwidth ]{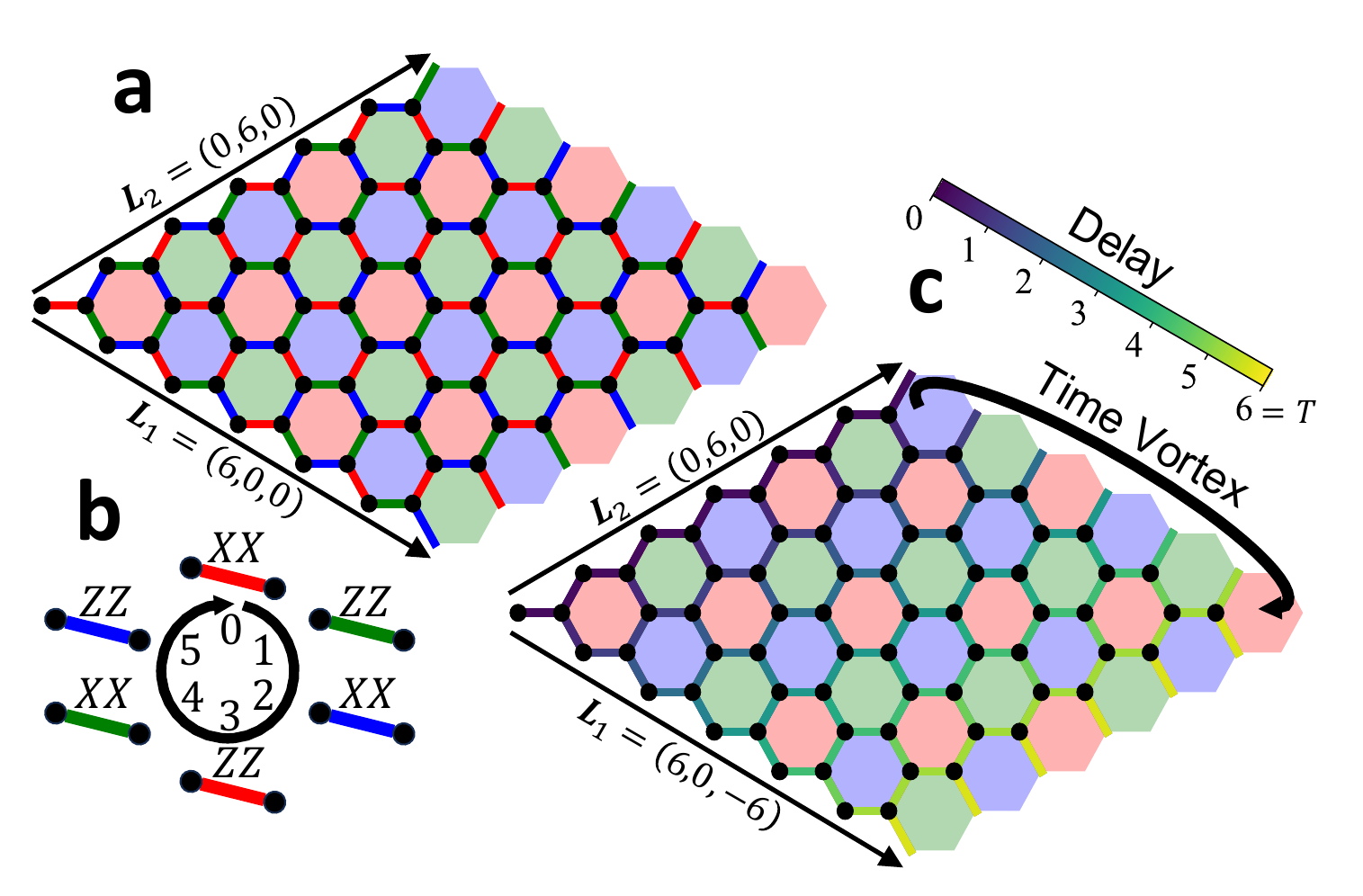}
    \caption{The Floquet color code with and without a time vortex. \textbf{(a)} The qubits of the Floquet color code are arranged on the vertices of a honeycomb lattice on a torus. The plaquettes are colored red, green, and blue and the bonds are colored according to the plaquettes at their endpoints. \textbf{(b)} The code is defined by a periodic sequence of 2-body Pauli $XX$ and $ZZ$ measurements on the bonds with period $T=6$. The time of each measurement is given by the schedule indicated on the black circle. \textbf{(c)} A time vortex is inserted along the $\mathbf{L}_1$ lattice vector by adding a spatially dependent delay to the time at which each measurement is applied within the cycle. The delay accumulated on a path going around the torus in the $\mathbf{L}_1$ direction is equal to the period $T$.
    }
    \label{fig: floquet color code}
\end{figure}

The order of the measurement sequence is defined through a coloring of the lattice. The plaquettes of the lattice are colored red, green, and blue, assigning adjacent plaquettes distinct colors (see Fig.~\ref{fig: floquet color code}). Each bond of the lattice is correspondingly colored according to the two plaquettes at its endpoints. Finally, the measurement sequence of the Floquet color code is $rx \rightarrow gz \rightarrow bx \rightarrow rz \rightarrow gx \rightarrow bz$, where $rx$ for example indicates that all red bonds are measured in the $X$ basis. Although the code is defined through the order of the measurements alone, we find it convenient to introduce a continuous time at which the measurements are performed, i.e. the measurements above are performed at integer times $i+kT$ where $i\in\{0,\dots,5\}$ for each of the six steps, $T=6$ is the period and $k\in\mathbb{Z}$ is the cycle number. The Floquet color code is illustrated in Fig.~\ref{fig: floquet color code}(a) and the measurement sequence is shown in Fig.~\ref{fig: floquet color code}(b).

Importantly, the system is not always in an eigenstate of all of the color code stabilizers. Each stabilizer anticommutes with some of the checks in one of the sets measured during the sequence. For example the $X$ stabilizer on a red plaquette anticommutes with each of the red bond measurements in the $Z$ basis touching that plaquette. It can be seen that at each step within the cycle, the state is an eigenstate of all of the color code stabilizers except those on the plaquettes of the color last measured, with the opposite Pauli label as that last measured. The instantaneous state is equivalent up to a local unitary transformation to the toric code on a triangular lattice. For instance, after measuring the red bonds in the $X$ basis, the two qubits on each red bond are projected into a two dimensional subspace. This subspace is equivalent to a single qubit on each red bond. The red $X$ stabilizers of the color code are equivalent to the vertex of the toric code, while the green and blue $Z$ plaquette stabilizers of the color code correspond to triangular plaquettes of the toric code.

On a torus, the Floquet color code encodes two logical qubits. The logical operators $\bar{X}_{1,2}$ and $\bar{Z}_{1,2}$ of these qubits are supported on homologically nontrivial cycles around the torus and are given by products of physical $X$ and $Z$ Paulis respectively. The exact form of these operators can be read off from the equivalent toric code. Importantly, these operators are non static, but rather evolve during the measurement sequence as one instance of the toric code is mapped to another.

\section{Inserting time vortices}
\label{sec: inserting time vortices}

A time vortex is a space-time defect one can add to a periodically driven system \cite{kishony2024topologicalexcitationstimevortices}. Starting from a translation invariant model such as the Floquet color code drawn in Fig.~\ref{fig: floquet color code}(a), a time vortex is inserted by adding a spatially dependent time delay which smoothly accumulates an integer multiple of full driving period $T$ upon encircling the defect core [see Fig.~\ref{fig: floquet color code}(c)]. If bond $b$ at physical location $\bm{r}_b$ is measured at time $t_b + kT, k\in\mathbb{Z}$ in the translation invariant model, it is measured at time $t_b - \tau(\bm{r}_b) + kT$ with the vortex. The position dependent delay $\tau(\bm{r})$ satisfies
\begin{align}
\oint \nabla\tau(\bm{r})\cdot \bm{dr} = nT,
\end{align}
where $n\in \mathbb{Z}$ in the number of time vortices inserted, and the integral is done on a path around the time vortex core. Here, it is convenient (but not necessary) to think of $\tau(\bm{r})$ as a continuous function of space, which is sampled at the center of the bonds $\bm{r}_b$ when determining the time of the measurements.  Fig.~\ref{fig: floquet color code}(c) shows the spatially dependent delay in the measurement sequence of the Floquet color code with a single time vortex going around one direction of the torus. 

We emphasize that the vortexed code is defined by precisely the {\it same set of checks} as the vortex-free code (requiring the same spatial connectivity). {\it The only difference is in the order} of the measurement sequence. Furthermore, if the gradient $\nabla\tau(\bm{r})$ is small enough, the {\it local} order of measurements is unaffected - the change is evident only on the {\it global} level. 

In the main text, we study the Floquet code embedded on a manifold with no boundaries (the torus). We insert time vortices around homologically non-trivial cycles around the torus and study the effect on the properties of the code. In Appendix \ref{app: planar}, we consider the implications for a planar architecture (with boundaries). 

The downside of introducing time vortices into the measurement sequence is that circuit depth becomes larger despite having the same number of operations. This overhead does not increase with the size of the code. 
In the presence of idling noise, this increases the logical error rate of the code. 
This is discussed in Appendix \ref{app: idle}, where we show that introducing time vortices may be beneficial even in the presence of idling noise, as long as it is small enough. 



\section{Optimal code with time vortices}
\label{sec:optimal}

In this section, we optimize the Floquet code graph-like distance
over the possible embeddings of the code on the torus and the numbers of time vortices in the two directions. 

\subsection{Graphlike distance}
\label{sec: distance}

As discussed in \cite{Anyons_color_code_Kesselring2024}, Floquet color code is matchable in the sense that independent single qubit Pauli errors and measurement errors trigger a non-trivial syndrome on up to two detection cells. This means that this error model can be represented on a ``matching graph'' such as the one illustrated in Fig.~\ref{fig: matching graph cartoon} with a vertex for each detector and an edge for each error. This structure allows for efficient decoding using methods such as minimum weight perfect matching.

For easy comparison with other literature, we focus primarily on the Majorana-inspired error model called ``EM3'' in \cite{Chao2020optimizationof, Gidney2022benchmarkingplanar}. This model assumes native 2-body Pauli measurements and includes correlated errors effecting measurement outcomes and the measured qubits. A detailed description of the model is given in Appendix \ref{app: error model}. Within this model, certain errors generate more than two detection events, but by decomposing these errors, efficient decoding is still possible.

In order to study the effect of insertion of time vortices in the Floquet color code we start by evaluating the effect on the graphlike distance of the code. The graphlike distance is the distance of the code if only those errors which trigger up to two detectors are considered. The true code distance may in general be smaller than the graphlike distance, but this metric can be computed efficiently and is a useful heuristic for optimizing the code as we do below. 

\begin{figure}
\centering\includegraphics[ width=0.8\columnwidth ]
{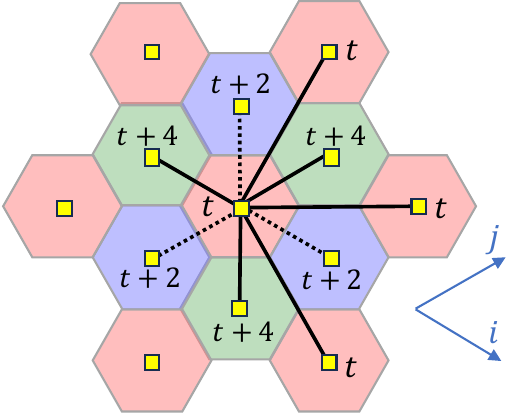}
\caption{The space-time matching graph of the Floquet color code for the EM3 error model. Vertices correspond to volume-like detector cells and edges correspond to errors that trigger pairs of detectors. The dotted edges belong to the set $E_2$ in Eq.~\eqref{eq: edges}.
The shortest path between distant points can be chosen such that it contains at most one edge of this set. 
}
\label{fig: matching graph actual}
\end{figure}

Since the Floquet color code is a CSS code, it suffices to analyze errors triggering only $X$ (or $Z$) type detectors. In the matching graph $G=(V,E)$, the $X$ type detectors of the Floquet color code are located at vertices of a space-time lattice $V=\{(i,j,t=2(i-j)+6k)\}$ with $i,j,k\in \mathbb{Z}$. Each of these corresponds to the product of two consecutive measurements of the Pauli $X^{\otimes 6}$ stabilizer at some plaquette of the original honeycomb lattice of qubits. Matchable errors are (undirected) edges $E=\{(v, v+w)| v\in V, w\in E_1\cup E_2 \cup E_3\}$, where
\begin{align}
    E_1 = &\{(-1,0,4), (0,1,4), (1,-1,4)\},\nonumber\\
    E_2 = &\{(1,0,2), (0,-1,2), (-1,1,2)\},\nonumber\\
    E_3 = &\{(1,1,0), (2,-1,0), (-1,2,0)\}.
    \label{eq: edges}
\end{align}
Each vertex has degree 18, i.e. each detector can be triggered by 18 different errors. This matching graph is drawn in Fig.~\ref{fig: matching graph actual}. We can freely choose how we embed our code on the torus by choosing two independent basis vectors 
\begin{equation}
\label{eq:L12}
\mathbf{L}_1=(a_1,b_1,-6n_1),\, \mathbf{L}_2=(a_2,b_2,-6n_2),    
\end{equation}
and identifying each point on the lattice with all points shifted by integer multiples of the basis vectors. The integers $n_1,n_2$ are the number of time vortices inserted around each direction of the torus. In order for the colors to be defined periodically, we set the vectors $\mathbf{L}_1,\mathbf{L}_2$ to be vectors of the superlattice which has one plaquette of each color in its unit cell - $\mathbf{L}_1=(c_1+3d_1,c_1,-6n_1)$, $\mathbf{L}_2=(c_2+3d_2,c_2,-6n_2)$.

The graphlike distance is equal to the length of the minimal length homologically non-trivial cycle in the matching graph. As noted in \cite{Sarkar2024graphbasedformalism} for the vortex-free case, when mapped back to the infinite lattice, such a path connects point $v$ with the point
$v + m_1\mathbf{L}_1 + m_2\mathbf{L}_2$ with at least one of the integers $m_1,m_2$ being odd. The graphlike code distance is therefore given by
\begin{align}
    D=\min_{\substack{m_1,m_2 \in \mathbb{Z}\\(m_1,m_2)\neq(0,0)}}\lVert m_1\mathbf{L}_1 + m_2\mathbf{L}_2\rVert_G,
    \label{eq: distance on torus}
\end{align}
where $\lVert w\rVert_G$ denotes the shortest distance in the graph $G$ from any vertex $v$ to the vertex $v+w$. Finding the distance $D$ is an instance of what is known as the shortest vector problem (SVP). In high dimensional spaces, this problem is extremely difficult. However in lower dimensions it becomes tractable. 

In order to express $\lVert \cdot\rVert_G$, we note that the sum of any two edges defined by $E_2$ in Eq.~\eqref{eq: edges} can be replaced with up to two edges from $E_1\cup E_3$. For instance, $2(1,0,2)=(1,-1,4)+(1,1,0)$ and $(0,-1,2)+(-1,1,2)=(-1,0,4)$. Therefore, any path on the graph with an even (odd) number of edges from $E_2$ can be replaced with a path of shorter or equal length with zero (one) edge from $E_2$. Omitting the edges from $E_2$ disconnects the graph into two connected components - vertices with time $t=0 \mod4$ and vertices with $t=2 \mod4$. A minimal length path between vertices in the same connected component can be found in the subgraph without edges from $E_2$. A minimal length path between vertex $v_1$ from the first connected component and $v_2$ from the second must have a single edge from $E_2$. The distance in this case is given by $\lVert v_2-v_1\rVert_G = 1+\min_{e\in E_2,s\in \pm 1}\lVert v_2-v_1+se\rVert_G$, where the distances to be computed on the right-hand side of the equality are within the same connected component.

The problem of computing distances in the graph $G$ has been reduced to computing distances only within the same connected component (omitting edges from $E_2$), i.e. $\lVert w=(i,j,t)\rVert_G$ with $t=0 \mod4$. We note that each of the three edges from the $E_3$ can be written as a sum of two edges from the $E_1$ taken with opposite signs, e.g. $(1,1,0)=(0,1,4)-(-1,0,4)$. Since the edges defined by $E_1$ span the entire connected component of the lattice (any two vertices can be connected along a path formed by these edges), we express any vector $w\in V$ in this basis
\begin{align}
    w = w_1(-1,0,4)+w_2(0,1,4)+w_3(1,-1,4),
\end{align}
where $w_1,w_2,w_3\in \mathbb{Z}$. Clearly, $|w_1|+|w_2|+|w_3|$ is an upper bound for $\lVert w\rVert_G$. If the three components $w_1,w_2,w_3$ have the same signs, this bound becomes tight - such a path traverses the lattice with the maximal velocity (of 4) in the time direction and any other path would take longer to reach the same time offset. If the signs of the components differ, pairs of edges with opposite signs can be replaced with a single edge from $E_3$ as described above. Therefore, we conclude that
\begin{align}
    \lVert w\rVert_G = \frac{1}{2}\left(|w_1|+|w_2|+|w_3|+|w_1+w_2+w_3|\right).
\end{align}

For example, the distance between points with space-like separation corresponds to $w_1+w_2+w_3=0$, in which case the shortest path is entirely along edges from $E_3$.

Now, in order to evaluate the distance by solving the SVP using Eq.~\eqref{eq: distance on torus}, the integers $m_1,m_2$ must be enumerated within the vicinity of $m_1=m_2=0$. For a choice of basis vectors $\mathbf{L}_1,\mathbf{L}_2$ which are large in magnitude and close to parallel, the region to be enumerated in order to minimize the distance will be very large. In order to simplify the computational task, we perform a basis reduction using the LLL algorithm \cite{Lenstra1982} to an equivalent basis with shorter and nearly orthogonal basis vectors. In this basis, it suffices to enumerate a small number of points $(m_1,m_2)$.

\subsection{Constraint on number of time vortices}
\label{sec: contraints on time vortices}

A given embedding of the code on the torus parametrized by $\mathbf{L}_1,\mathbf{L}_2$ (including the specification of the number of time vortices $n_1,n_2$) is allowed if the local order of measurements is the same as in the vortex-free code, up to a spatially dependent delay. In other words, when looking at the measurements of the bonds incident to any qubit, the order must be $rx \rightarrow gz \rightarrow bx \rightarrow rz \rightarrow gx \rightarrow bz$ up to a cyclic permutation. This condition may break down if the gradient of the delay is too large.

The delay induced by the time vortices between bonds separated by a spatial offset $(i, j)$ in the coordinate system illustrated in Fig.~\ref{fig: matching graph actual} is given by
\begin{align}
    \delta t = 6\frac{n_1\left(b_2i-a_2 j\right) + n_2\left(a_1j-b_1i\right)}{a_1b_2-a_2b_1},
\end{align}
where $a_{1,2}$, $b_{1,2}$ $n_{1,2}$ define the basis vectors in Eq.~\eqref{eq:L12}. 
In the vortex-free code, the spatial offset between two bonds incident to the same qubit, where the first is measured one time unit before the second within the cycle is given by one of the following three vectors:
\begin{align}
    (i_1,j_1) &= (\tfrac{1}{2},0), \nonumber\\
    (i_2,j_2) &= (-\tfrac{1}{2},\tfrac{1}{2}), \nonumber\\
    (i_3,j_3) &= (0,-\tfrac{1}{2}).
\end{align}

In order for the local measurement sequence to remain unaffected by the vortices, each of the three corresponding delays $\delta t_1, \delta t_2, \delta t_3$ must be bounded between $-1$ and $+5$. This results in the following constraints on the allowed number of vortices:
\begin{align}
\label{eq:conditions}
    -1<\delta t_1=3\frac{n_1\left(b_2\right) + n_2\left(-b_1\right)}{a_1b_2-a_2b_1}<5,\nonumber\\
    -1<\delta t_2=3\frac{n_1\left(-b_2-a_2 \right) + n_2\left(a_1+b_1\right)}{a_1b_2-a_2b_1}<5,\nonumber\\
    -1<\delta t_3=3\frac{n_1\left(a_2 \right) + n_2\left(-a_1\right)}{a_1b_2-a_2b_1}<5.\nonumber\\
\end{align}
A number of vortices proportional to the linear
size of the system can be inserted around each direction of the torus without violating Eq.~\eqref{eq:conditions}. 

\subsection{Optimal embedding on torus with and without vortices}
\label{sec: optimal R}

Varying the choice of embedding of the code on the torus by changing $\mathbf{L}_1,\mathbf{L}_2$ yields codes with different parameters $[\![ N, K, D]\!]$. For any valid choice, the number of encoded logical qubits is equal to $K=2$. The number of physical qubits is given by
\begin{align}
    N = 2\left|a_1b_2-a_2b_1\right|,
\end{align}
and the distance $D$ is calculated using Eq.~\eqref{eq: distance on torus}. The figure of merit for performance of these codes is $R=N/D^2$, which remains constant when the code is simply scaled by a constant factor ($\mathbf{L}_i\rightarrow\alpha \mathbf{L}_i$). In the thermodynamic limit, the best possible value of $R$ must saturate to some constant according to the BPT bound \cite{Bravyi2010BPTbound}.

We perform an exhaustive search over choices of $\mathbf{L}_1,\mathbf{L}_2$ with bounded $N<1000$ qubits in order to find the best performance achievable both with and without time vortices. All distinct optimal configurations found are presented in Appendix \ref{app: optimal}.

\begin{figure}
\centering\includegraphics[ width=1\columnwidth ]
{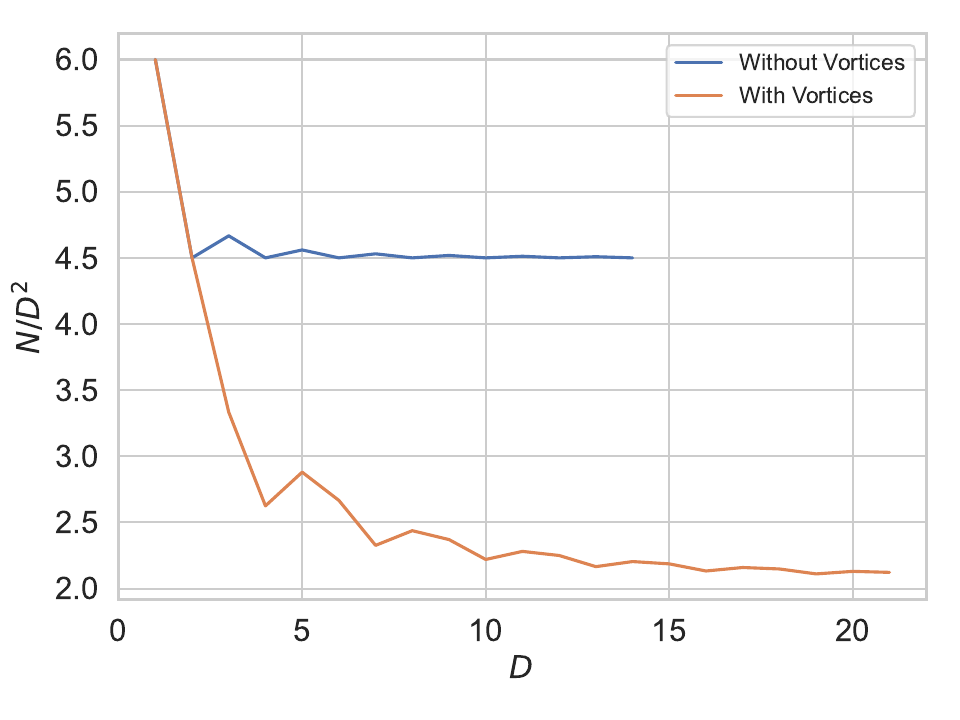}
\caption{Lowest value of $R=N/D^2$ reached by choosing the optimal embedding of the Floquet color code on the torus with and without time vortices as a function of the code distance $D$.
}
\label{fig: R vs. D}
\end{figure}

We find that the best configuration in the vortex-free case is given by $\mathbf{L}_1=(3x,0,0)$ and $\mathbf{L}_2=(0,3x,0)$ which yields a distance of $D=2x$ and $N=18x^2$ physical qubits, i.e. $R=4.5$. This is the same configuration as the one suggested by \cite{bombin2007optimalresources} for the static (not Floquet) color code. The best configuration we find with vortices is $\mathbf{L}_1=(19x, 1x, 36x)$, $\mathbf{L}_2=(1x, -20x, -72x)$. For this choice, $D=19x$, $N=762x^2$ and $R\approx 2.11$. We note that the optimal configuration in the presence of time vortices is not a symmetric one, as opposed to the vortex-free case. Fig.~\ref{fig: R vs. D} shows the best value of $R$ found for each code distance $D$, with and without time vortices. At large code distances inserting time vortices allows for a reduction by a factor of at least $2.13$ in the number of physical qubits required, and already at a modest distance of $4$, the number of qubits can be reduced by a factor of $1.71$.

\section{Simulations}
\label{sec: simulations}

In this section, we perform Monte Carlo simulations using  using the open-source tool Stim \cite{Gidney2021stimfaststabilizer} in order compare the performance of the vortexed version of the Floquet color code with the vortex-free one under the circuit-level noise model EM3. Decoding is done with a minimum-weight perfect matching decoder, using PyMatching \cite{higgott2023sparse}. The Python code used to generate the results presented in this section is available online at \url{https://github.com/kishonyWIS/measurementTimeVortex}.

We perform quantum memory experiment simulations in which a known logical state is initialized, then a certain number of noisy error correction rounds are performed, and finally the logical state is read out projectively. For simplicity, we assume the initialization and measurement stages are noise-free as done in \cite{Anyons_color_code_Kesselring2024}. The number of error correction rounds is chosen to be proportional to the graphlike code distance, as the duration of logical operations is expected to scale in this manner with the code distance. The initial logical state is an eigenstate of one horizontal and one vertical logical operator ($\bar{X}_{1,2}=1$), such that logical error rates for both can be assessed simultaneously. For simplicity, only detectors obtained from Pauli $X$ checks were input for decoding (as done in \cite{Anyons_color_code_Kesselring2024}), although in principle, performance may be improved by considering all detectors within the same decoding problem.

We simulate the best vortexed and vortex-free codes found in Sec.~\ref{sec: optimal R} with up to $100$ qubits. For certain code distances, we find more than one distinct configuration with the optimal number of physical qubits in the vortexed case (see Appendix \ref{app: optimal}). Here we present the ones we find to perform best in the numerics: $\mathbf{L}_1=(4, 4, -18), \mathbf{L}_2=(6, -3, -12)$ at distance $D=5$ and $\mathbf{L}_1=(1, 7, -12), \mathbf{L}_2=(7, 1, 6)$ at $D=6$.

\begin{figure}
\centering\includegraphics[ width=1\columnwidth ]
{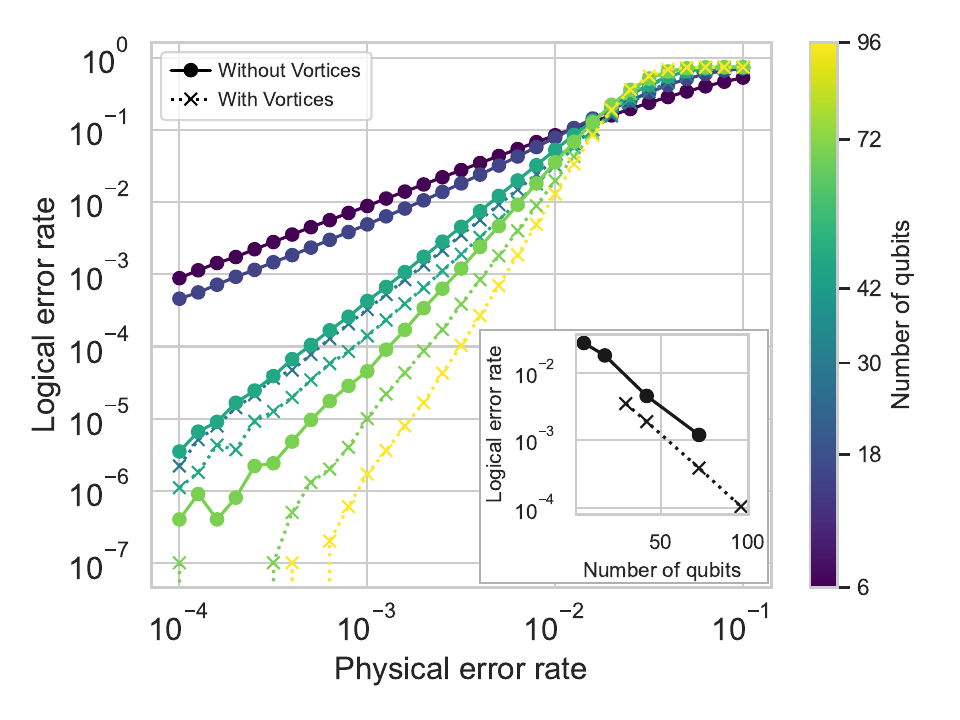}
\caption{The logical error rate vs. the physical error rate for optimal embeddings of the Floquet color code on the torus with different code distances (determined by the basis vectors $\mathbf{L}_1,\mathbf{L}_2$). Configurations without time vortices are drawn with circular markers and solid lines and those with vortices are drawn with x markers and dotted lines. The color indicates the number of physical qubits in the code block. The inset shows the logical error rate vs. the number of qubits at a fixed physical error rate $p=10^{-2.5}$.
}
\label{fig: threshold}
\end{figure}

Fig.~\ref{fig: threshold} presents the logical error rate vs. the physical error rate for each of these codes. The inset shows the logical error rate as a function of the number of physical qubits at a fixed physical error rate. We find that the threshold of both code families is roughly the same, about $1.6\%-2\%$, which is consistent with \cite{Gidney2022benchmarkingplanar}. While the time vortex does not improve the threshold of the code, it dramatically reduces the number of physical qubits needed in order to reach a given logical error rate. With an equal number of physical qubits, the vortexed code outperforms the vortex-free code for any physical error rate below the threshold. For example, the smallest vortexed code presented (given by $\mathbf{L}_1=(3,0,-6), \mathbf{L}_2=(1,-5,0)$) with $30$ qubits even slightly outperforms the vortex-free code with the same distance (of $3$) and $42$ qubits (given by $\mathbf{L}_1=(4,1,0), \mathbf{L}_2=(1,-5,0)$).


\section{The toric code}
\label{sec: toric code}

In this section, we briefly consider the application of our time vortex construction to the setting of the well known toric code \cite{Kitaev1997toriccode, kitaev2003toriccode} (referred to as the surface code in the planar setting \cite{dennis2002surfacecode, Fowler2012surfacecode}). Despite discussing the Floquet color code in detail above, we stress that the time vortex construction does not rely on the existence of non-commuting checks within measurement sequence, 
as long as the stabilizers are measured by first entangling the data qubits with auxiliary qubits and then measuring the latter, 
as exemplified here for the toric code.

\begin{figure}[H]
    \centering
    \includegraphics[width=1\columnwidth]{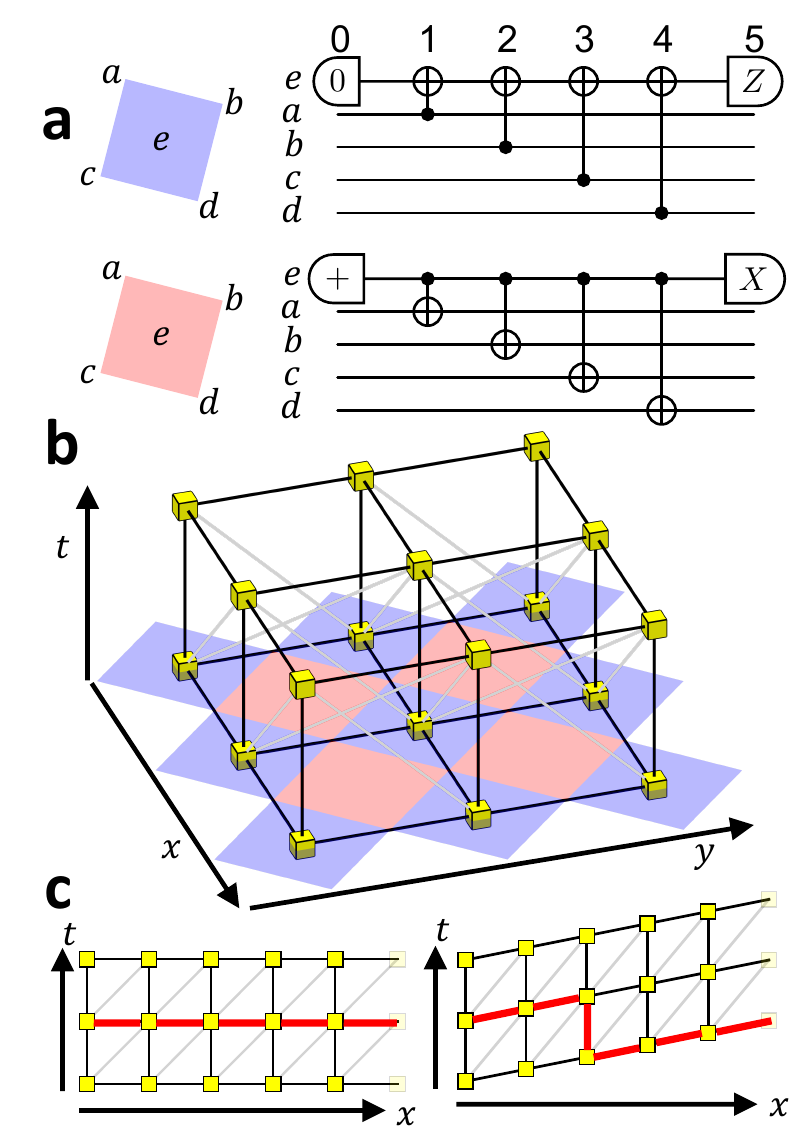}
    \caption{Error propagation in the toric code. \textbf{(a)} Syndrome measurement is performed using an auxiliary qubit at the center of each plaquette. 
    $Z$ stabilizers (blue) and $X$ stabilizers (red)
    are measured in parallel according to the circuits illustrated. \textbf{(b)} The matching graph of $Z$ type detectors. Black edges correspond to phenomenological errors: time-like edges are measurement errors and space-like edges are $X$ errors between syndrome measurement rounds. The diagonal gray edges are circuit-level errors occurring at intermediate times during the syndrome measurement circuit. Diagonals between detectors which are neighboring in space can be caused for example by an $X$ error on the data qubit shared by the two stabilizers between step $2$ and $3$ of the circuit in \textbf{(a)}. Diagonals between second-neighbors can be caused by an $X$ error on the auxiliary qubit at the center of a red plaquette between step $2$ and $3$, which propagates into the data qubits (this error is known as a hook error). \textbf{(c)} Inserting time vortices in the positive $x$ direction increases the code distance, but negative vortices do not (due to the gray edges).
    }
    \label{fig: matching surface code}
\end{figure}

The toric code is a CSS code defined on a square lattice with qubits on the vertices. The plaquettes of the lattice are colored red and blue in a checkerboard pattern where red plaquettes host $X$ type stabilizers and blue plaquettes host $Z$ type stabilizers of weight $4$ (see Fig.~\ref{fig: matching surface code}). 
The stabilizers are mutually commuting. This implies that the order in which they are measured has no consequence in terms of the code space, but the details of the syndrome measurement circuit (which uses auxiliary qubits and two-qubits Clifford gates) may have consequences in terms of error propagation, depending on the error model \cite{dennis2002surfacecode, fowler2009high_threshold, Fowler2012surfacecode,orourke2024comparepairrotatedvs}.

Fig.~\ref{fig: matching surface code}(a) presents a particular choice of circuit used for syndrome extraction and Fig.~\ref{fig: matching surface code}(b) shows the corresponding matching graph of the $Z$ type detectors for phenomenological and circuit-level noise (the matching graph for the $X$ type detectors is identical). For the simple phenomenological noise model, the matching graph is a cubic lattice with edges between nearest neighbors. Inserting a time vortex around either direction of the torus with a positive or negative sign increases the code distance with respect to errors in the same direction [analogously to Fig.~\ref{fig: matching graph cartoon}(b)]. 

Including circuit-level noise adds diagonal edges to the matching graph. In this case the code distance in the $x$ direction of the torus can be increased by inserting a time vortex with the correct sign such that a minimal weight undetectable error does not include diagonal edges [Fig.~\ref{fig: matching surface code}(c)].

The toric code is a static code in the sense that its stabilizers and code space remain fixed at each round of measurements. Nonetheless, the circuit used to propagate the value of these stabilizers onto the auxiliary qubits results in non-trivial dynamics at intermediate times. At all intermediate times the auxiliary qubits and data qubits become entangled while at the end of each round they are decoupled. When a time vortex is inserted, there is no longer a plane in space-time on which the auxiliary qubits and data qubits are decoupled. From this perspective, the time-vortex measurement dynamics puts the auxiliary qubits to dual use: they intermittently become part of the code’s support during the entangling steps, and are then measured out to yield the syndrome information. For this reason, although the vortex-free code suffers from an undetectable nontrivial error of weight equal to the linear system size $L$ acting at equal time, for the vortexed code no such error exists.

To provide more intuition for this behavior, Appendix \ref{app: rep code with ancillas} presents the circuit for syndrome measurement in the repetition code using auxiliary qubits and two-qubit Clifford gates as done here for the toric code. This can be obtained simply by reducing the width of the toric code in Fig.~\ref{fig: matching surface code} in the $y$ direction to a single qubit. The matching graph for this circuit with and without a time vortex is therefore exactly as shown in Fig.~\ref{fig: matching surface code}(c) for the toric code as viewed from the side.

As in the Floquet color code, simultaneously optimizing the embedding of the toric code on a torus and the number of vortices inserted in each direction will lead to a reduction in the number of qubits for a given distance by a constant factor when compared with the optimal vortex-free code. We leave such an optimization for future work. 

\section{Summary}

To summarize, in this work, we suggest improving the performance of topological codes by introducing time vortices into the measurement sequence defining them. We demonstrate the power of this construction within the framework of the Floquet color code, which is considered to be the state-of-the-art for hardware with native 2-body Pauli measurements on a two dimensional lattice. 

Our construction also combines spatial ``twists'' \cite{bombin2007optimalresources,kovalev2011additivequantumcodes, breuckmann2021balancedproductcodes, Sarkar2024graphbasedformalism}, in which the geometry of the lattice is sheared in the two spatial directions. While the best performance is achieved by combining twists in space and time, we note that our time vortex construction requires no spatial deformation of the lattice if this is not possible on the hardware; in this case, the connectivity required for the error correction circuit is precisely the same as that required for the vortex-free code. 

It would be interesting to consider the possibility of generalizing the time vortex construction for code families with better asymptotic parameters such as hypergraph product codes \cite{Tillich_2014, bravyi2013homologicalproductcodes}, or even ``good'' low density parity check (LDPC) codes \cite{panteleev2022asymptoticallygoodquantumlocally, leverrier2022quantumtannercodes, dinur2022goodquantumldpccodes} with $K/N=O(1)$ and $D/N=O(1)$.

\acknowledgements

We thank Yaar Vituri, Yarden Sheffer, Shoham Jacoby, Peter-Jan H. S. Derks, and Julio C. Magdalena de la Fuente for useful discussions.
This work was supported by CRC 183 of the Deutsche Forschungsgemeinschaft (project number 277101999, subproject A01), a research grant from the Estate of Gerald Alexander, the CHE/PBC Doctoral Fellowships in Quantum Science and Technology 2024, and ISF-MAFAT Quantum Science and Technology Grant no. 2478/24.

\bibliographystyle{unsrtnat}
\bibliography{literature.bib}

\clearpage
\appendix

\section{The repetition code without native two-body measurements}
\label{app: rep code with ancillas}

\begin{figure}[H]
\centering\includegraphics[ width=0.9\columnwidth ]
{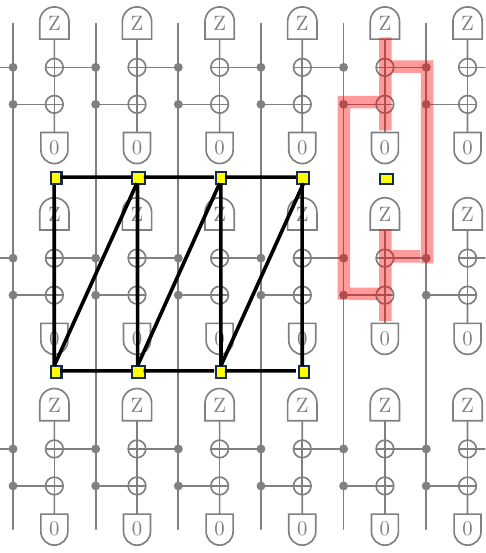}
\caption{Measurement of the repetition code stabilizers $Z_iZ_{i+1}$ with periodic boundary conditions using auxiliary qubits and Clifford gates. The circuit is drawn in gray, with time evolving from bottom to top. The red outline indicates the detecting region of one of the detectors given by the product two consecutive measurements of the same stabilizer - the locations of all errors which can trigger the detector. These detecting regions are tiled in space and time as indicated by the yellow squares. Edges connecting these detectors correspond to all possible circuit-level bit flip errors; each error triggers a pair of detectors. Space-like edges correspond to errors on data qubits between rounds of stabilizer measurements. Time-like edges correspond to measurement errors (flipped auxiliary qubits). A diagonal edge arises from a data qubit flipping at an intermediate time during the circuit. The resulting matching graph is identical to that shown in Fig.~\ref{fig: matching surface code}(c).        
}
\label{fig: rep code with ancillas}
\end{figure}

We return to the repetition code which was used as a toy example to illustrate the time vortex construction in Sec.~\ref{sec: intro}. Here, we consider the case in which auxiliary qubits and two-qubit Clifford gates are used to measure the stabilizers of this code in place of native two-body Pauli measurements. Fig.~\ref{fig: rep code with ancillas} shows the circuit for repeatedly measuring the stabilizers of this code (defined with periodic boundary conditions). This circuit can in fact be obtained as a special case of the circuit used to measure the stabilizers of the toric code in Fig.~\ref{fig: matching surface code} by taking the width of the system in the $y$ dimension to a single qubit; the repetition code is a one-dimensional toric code.

The matching graph for circuit-level bit flip errors is overlaid on top of the circuit in Fig.~\ref{fig: rep code with ancillas}. The detecting region or Pauli web of one of the detectors is highlighted \cite{McEwen2023relaxinghardware, Bombin2024unifyingflavorsof} - this is the set of all circuit locations at which an error can trigger this detector. This allows to visually identify the effect of any bit flip error on the nearby detectors and to confirm the structure of the matching graph drawn. Uncoincidentally, this matching graph is exactly the same as the projected view of the toric code matching graph drawn in Fig.~\ref{fig: matching surface code}(c). As explained for the toric code in Sec.~\ref{sec: toric code}, time vortices inserted in one direction increase the distance of this code, while in the other direction, the diagonal edges in the matching graph result in the distance remaining unchanged.

It may seem surprising that the distance of the repetition code with respect to circuit-level bit flip errors can be made larger than the number of data qubits $N$; it would seem that an equal-time error on all data qubits would be an undetectable non-trivial logical error. However, as in the case of the toric code, while the repetition code is static in the absence of time vortices when viewed {\it stroboscopically} at the end of each round of stabilizer measurements, the syndrome measurement circuit entangles the data qubits with the auxiliaries at intermediate times. In fact, the instantaneous code space after the first round of entangling gates is that of a repetition code on all $2N$ qubits. Once vortices are introduced, some subset of auxiliary qubits is entangled with the data qubits at every instant throughout the code's evolution under the circuit.

\section{Circuit depth and idle qubits}
\label{app: idle}

When time vortices are introduced into the measurement sequence of a topological code, the circuit depth is increased. In the vortex-free Floquet color code for instance, the measurements are ``tightly packed'' in the sense that every qubit participates in some measurement during every layer of the circuit. This is not the case in the vortexed code - on average an extensive number of qubits are idle during each layer of measurements. 

The depth of the circuit increases by a factor $1+\alpha x$ where 
$\alpha=n/L$, $n$ is the number of time vortices, $L$ is the linear system size, and $x$ is a positive constant. The code distance increases by a corresponding factor $D=D_0(1+\alpha y)$ with another positive constant $y$. The spatial gradient of the time delay induced by the time vortices $\alpha$ is bounded by a system size independent constant; the number of time vortices is bounded by the linear system size. Therefore, the overhead in circuit depth compared with the vortex-free code is only a constant factor. If the physical qubits are subject to idling noise, the probability of an error per qubit per cycle increases as $p\rightarrow (1+\alpha z)p$ with some constant $z$. Overall, the logical error rate scales as
\begin{align}
    \bar{p} \propto \left[p(1+\alpha z)\right]^{\frac{1}{2}D_0(1+\alpha y)}.
\end{align}
To leading order in $\alpha$, inserting time vortices leads to an exponential reduction in the logical error rate as long as $y\log(p)+z<0$. On many devices, idling errors are significantly smaller than gate errors such that $z$ is very small.

Furthermore, it may be beneficial to incorporate repeated measurements (as done in Ref.~\cite{PhysRevX.11.031039} for subsystem codes) on bonds connecting qubits that would otherwise be idle within a layer of the circuit of measurements of a vortexed code. This has been shown to improve performance in some cases even in the vortex-free case where repeating measurements comes at the expense of an extended cycle duration which would otherwise be compact. 
We leave the exploration of these possibilities for future work.

\section{Time vortices around punctures in a planar architecture}
\label{app: planar}

\begin{figure}
\centering\includegraphics[ width=0.7\columnwidth ]
{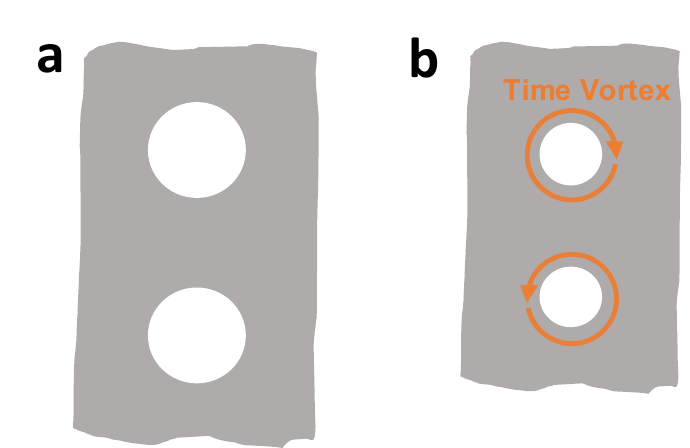}
\caption{A topological Floquet code embedded on a planar lattice of physical qubits. \textbf{(a)} Logical qubits can be encoded by making punctures inside which the measurement sequence is turned off. The code distance is determined by the circumference of the punctures and by their spatial separation. \textbf{(b)} By inserting time vortices around each puncture, the circumference of the punctures can be reduced while maintaining the same code distance, leading to a reduction in the number of physical qubits per logical qubit.
}
\label{fig: planar}
\end{figure}

Many quantum hardware platforms offer connectivity between physical qubits which is restricted to a 2D planar lattice. Embedding topological codes on a plane (instead of a torus) on such platforms requires a construction of boundaries. Simple constructions of boundaries for Floquet codes have been suggested \cite{Haah2022boundarieshoneycomb, Gidney2022benchmarkingplanar}. In order to increase the number of encoded logical qubits on the plane, more varied boundaries must be created. This can be done by partitioning the plane into disjoint patches, each hosting $O(1)$ logical qubits \cite{Horsman_2012}, or by creating punctures (regions in which the measurement sequence is disabled) within the fully connected plane \cite{Fowler2012surfacecode}. In the latter option [illustrated in Fig.~\ref{fig: planar}(a)], logical operators have support on cycles going around these punctures, and on paths connecting one puncture to another. 

Within this setup, we suggest inserting time vortices around each puncture. This should allow one to make the punctures smaller while maintaining the code distance, reducing the footprint of each logical qubit on the plane as illustrated in Fig.~\ref{fig: planar}(b).

\section{The EM3 error model}
\label{app: error model}

Throughout this work, we focus primarily on the EM3 error model \cite{Chao2020optimizationof, Gidney2022benchmarkingplanar}. We consider errors only in the bulk of the circuit for our Monte Carlo simulations in Sec.~\ref{sec: simulations}, and perform initialization and logical readout with noise-free circuits.

The EM3 noise model attempts to capture errors expected from a direct hardware implementation of two-qubit parity measurements. It is defined by replacing each measurement of a two-qubit Pauli by the following sequence with probability $p$, and performing the ideal measurement with probability $1-p$ ($p$ is the physical error rate on the $x$ axis of Fig.~\ref{fig: threshold}).

\begin{enumerate}
    \item Choose a two-qubit Pauli error uniformly randomly from the set $\{I, X, Y, Z\}^{\otimes 2}$ and apply it to the target qubits of the measurement.
    \item Perform the ideal measurement.
    \item Choose a ``measurement error'' randomly from the set $\{\text{flip}, \text{no flip}\}$ with equal probability. If $\text{flip}$ is selected, flip the result of the measurement outcome.
\end{enumerate}

Within the Floquet color code, each of these errors has a different effect on the detection cells incident to the bond being measured. Since the code is of the CSS family, we consider only the effect on the $X$ type detectors. A single qubit Pauli $Z$ (or $Y$) error triggers the two detectors incident to the qubit which are active at the time of the error (at any time there are active detectors on two of the three neighboring plaquettes). This is a matchable error which is represented by an edge defined by the set $E_2$ in Eq.~\eqref{eq: edges}. A measurement error triggers a detector on each of the two plaquettes that share the bond being measured. This is a matchable error as well, and corresponds to an edge from the set $E_1$.

Within the EM3 error model, any combination of Pauli errors on the two measured qubits and a measurement error may occur with probability $O(p)$. Each such combination triggers some subset of detectors on the four plaquettes that share at least one site with the measured bond. There are two subsets of these detectors which can be triggered by such a correlated error other than the subsets triggered by the individual errors discussed above. The first subset contains the detectors on the two plaquettes at the endpoints of the measured bond. This is a matchable error and is captured by an edge from the set $E_3$. This case includes for instance a $Z\times Z\times \text{no flip}$ error occurring during an $XX$ check measurement. The second subset contains detectors on all four plaquettes adjacent to the measured bond, arising for example from a $Z\times Z\times \text{flip}$ error during an $XX$ measurement. These errors are not matchable, and are therefore omitted from the discussion in Sec.~\ref{sec: distance}, but are accounted for in our Monte Carlo simulations in Sec.~\ref{sec: simulations}.

\section{All optimal configurations on the torus}
\label{app: optimal}

\begin{table*}[t]
\centering
\caption{Optimal embedding of the Floquet color code on the torus with and without time vortices.}
\small
\begin{tabular}{c|ccc|ccc}
\toprule
\hline
\hline
& \multicolumn{3}{c|}{\textbf{Without Vortices}} & \multicolumn{3}{c}{\textbf{With Vortices}} \\
\midrule
\hline
\textbf{$D$} & \textbf{$N$} & \textbf{$\mathbf{L}_1$} & \textbf{$\mathbf{L}_2$} & \textbf{$N$} & \textbf{$\mathbf{L}_1$} & \textbf{$\mathbf{L}_2$} \\
\midrule
1 & 6 & (1, 1, 0) & (2, -1, 0) & 6 & (1, 1, 0) & (2, -1, 0) \\
2 & 18 & (3, 0, 0) & (0, 3, 0) & 18 & (3, 0, 0) & (0, 3, 0) \\
3 & 42 & (4, 1, 0) & (1, -5, 0) & 30 & (3, 0, -6) & (1, -5, 0) \\
4 & 72 & (0, 6, 0) & (6, 0, 0) & 42 & (1, 4, 12) & (5, -1, 6) \\
5 & 114 & (7, 1, 0) & (1, -8, 0) & 72 & (3, 3, 12) & (5, -7, -6) \\
 & - & - & - & 72 & (6, 0, 6) & (0, 6, -6) \\
 & - & - & - & 72 & (4, 4, -18) & (6, -3, -12) \\
 & - & - & - & 72 & (0, 6, -6) & (6, 3, -18) \\
6 & 162 & (0, 9, 0) & (9, 0, 0) & 96 & (1, 7, -12) & (7, 1, 6) \\
 & - & - & - & 96 & (2, 5, -12) & (8, -4, -18) \\
7 & 222 & (1, 10, 0) & (11, -1, 0) & 114 & (1, 7, -12) & (8, -1, -24) \\
8 & 288 & (0, 12, 0) & (12, 0, 0) & 156 & (3, 6, -18) & (11, -4, -30) \\
9 & 366 & (13, 1, 0) & (1, -14, 0) & 192 & (3, 9, -12) & (10, -2, 18) \\
10 & 450 & (15, 0, 0) & (0, 15, 0) & 222 & (1, 10, 36) & (11, -1, 18) \\
11 & 546 & (16, 1, 0) & (1, -17, 0) & 276 & (3, 9, -24) & (14, -4, -42) \\
12 & 648 & (18, 0, 0) & (0, 18, 0) & 324 & (2, 11, -24) & (14, -4, -42) \\
13 & 762 & (1, 19, 0) & (20, -1, 0) & 366 & (13, 1, 24) & (1, -14, -48) \\
14 & 882 & (21, 0, 0) & (0, 21, 0) & 432 & (2, 14, -24) & (15, -3, 30) \\
15 & - & - & - & 492 & (4, 13, 54) & (18, -3, 24) \\
16 & - & - & - & 546 & (16, 1, -60) & (1, -17, 30) \\
17 & - & - & - & 624 & (15, 3, 36) & (4, -20, -66) \\
18 & - & - & - & 696 & (3, 18, -30) & (19, -2, 36) \\
19 & - & - & - & 762 & (19, 1, 36) & (1, -20, -72) \\
20 & - & - & - & 852 & (20, 2, 36) & (3, -21, 42) \\
21 & - & - & - & 936 & (20, 2, 42) & (4, -23, -78) \\

\hline
\hline
\bottomrule
\end{tabular}
\label{tab: optimal}
\end{table*}

Here, we present all of the distinct optimal embeddings of the Floquet color code on the torus both with and without time vortices found by an exhaustive search (see Sec.~\ref{sec: optimal R}). We say that two configurations $\mathbf{L}_1^a, \mathbf{L}_2^a$ and $\mathbf{L}_1^b, \mathbf{L}_2^b$ are equivalent if they are related by the point group symmetry of the code combined with a determinant $\pm 1$ linear transformation. A linear transformation of the basis vectors defined by a determinant $\pm1$ matrix over the integers yields a different basis for the same integer valued lattice. The point group symmetries of the Floquet color code model are generated by a reflection $(i,j,t)\rightarrow (-i,i+j,t)$ and the combination of a $60$ degree rotation with time reversal $(i,j,t)\rightarrow (-j,i+j,-t)$ (see Fig.~\ref{fig: matching graph actual}).

We exhaustively search over all possible configurations up to $N=1000$ qubits. For each code distance, we list all distinct instances with the minimal number of qubits in Table \ref{tab: optimal}, both for the vortex-free case and  with arbitrary time vortices. For $D=5$ we find $4$ distinct configurations with the same number of qubits $N=72$ with vortices, and for $D=6$ there are two distinct configurations with $N=96$. Despite having the same distance, these configurations perform differently in the presence of circuit-level noise due to different combinatorial possibilities for errors of the same weight. However, the distance controls the behavior at low physical error rates or when the code is scaled to large sizes. In Fig.~\ref{fig: threshold} we present the configuration with the best performance out of these options.

\end{document}